\documentclass[journal]{IEEEtran}
\usepackage{graphicx}
\usepackage{booktabs}
\usepackage{subfigure}
\usepackage{overpic}
\usepackage{amsmath}
\usepackage{flushend}
\usepackage{amssymb}
\usepackage{multirow}
\usepackage{makecell}
\usepackage{fontawesome}
\usepackage{color, soul}
\soulregister\cite7

\usepackage[ruled]{algorithm2e}
\usepackage{hyperref}
\hypersetup{
	colorlinks=true,
	linkcolor=blue,
	urlcolor=blue,
	citecolor=blue}
\usepackage{colortbl}
\usepackage[dvipsnames]{xcolor}
\definecolor{Bg}{HTML}{e0f1ff}

\begin{document}

\title{Dynamic Frequency Feature Fusion Network for Multi-Source Remote Sensing Data Classification}
\author{
    Yikang Zhao, 
    Feng Gao, \emph{Member, IEEE},
    Xuepeng Jin,
    Junyu Dong, \emph{Member, IEEE},  \\
    and Qian Du, \emph{Fellow, IEEE}
\thanks{This work was supported in part by the National Science and Technology Major Project under Grant 2022ZD0117201, and in part by the Natural Science Foundation of Shandong Province under Grant ZR2024MF020. (\textit{Corresponding author: Feng Gao})

Yikang Zhao, Feng Gao, Xuepeng Jin, and Junyu Dong are with the State Key Laboratory of Physical Oceanography, Ocean University of China, Qingdao 266100, China. (email: gaofeng@ouc.edu.cn)

Qian Du is with the Department of Electrical and Computer Engineering, Mississippi State University, Starkville, MS 39762 USA.}}

\markboth{IEEE GEOSCIENCE AND REMOTE SENSING LETTERS}%
{Shell}

\maketitle

\begin{abstract}

Multi-source data classification is a critical yet challenging task for remote sensing image interpretation. Existing methods lack adaptability to diverse land cover types when modeling frequency domain features. To this end, we propose a Dynamic Frequency Feature Fusion Network (DFFNet) for hyperspectral image (HSI) and Synthetic Aperture Radar (SAR) / Light Detection and Ranging (LiDAR) data joint classification. Specifically, we design a dynamic filter block to dynamically learn the filter kernels in the frequency domain by aggregating the input features. The frequency contextual knowledge is injected into frequency filter kernels. Additionally, we propose spectral-spatial adaptive fusion block for cross-modal feature fusion. It enhances the spectral and spatial attention weight interactions via channel shuffle operation, thereby providing comprehensive cross-modal feature fusion. Experiments on two benchmark datasets show that our DFFNet outperforms state-of-the-art methods in multi-source data classification. The codes will be made publicly available at \url{https://github.com/oucailab/DFFNet}.

\end{abstract}

\begin{IEEEkeywords}
Multi-source data classification; 
Hyperspectral image;
Synthetic aperture radar;
Light detection and ranging;
Dynamic frequency feature fusion.
\end{IEEEkeywords}

\IEEEpeerreviewmaketitle

\section{Introduction}

\IEEEPARstart{I}{n} recent years, with the rapid development of earth observation systems and satellite sensors, a vast amount of remote sensing images has been collected. Among these datasets, hyperspectral images (HSIs) have garnered significant attention due to their rich spectral information and have been widely utilized in land cover classification \cite{luo24tgrs}. Nevertheless, HSIs are highly vulnerable to atmospheric and lighting conditions. Specifically, the presence of cloud cover or inadequate illumination can degrade data quality, thus compromising classification performance. To tackle this challenge, Light Detection and Ranging (LiDAR) and Synthetic Aperture Radar (SAR) data are often leveraged to provide complementary information.  LiDAR delivers high-resolution 3D point clouds for structural analysis and offer elevation and structural details of ground objects, while SAR data are capable of imaging under all weather conditions and during both day and night. The integration of HSI with SAR/LiDAR data capitalizes on the complementary advantages of multi-source remote sensing data, enhancing the accuracy and robustness of land cover classification \cite{gyh24tip}. Therefore, this study focuses on the joint classification of HSI and SAR/LiDAR data.

Numerous deep learning-based methodologies for the joint classification of multi-source remote sensing data have been proposed. These methods mainly consist of three parts: multi-source feature extraction, cross-modal feature fusion, and classification \cite{nk25grsl}. Convolutional neural networks (CNNs) and Transformers are commonly used for multi-source feature extraction. In terms of cross-modal feature fusion, techniques such as dual-channel spatial attention \cite{ghm23tgrs}, spatial-domain geometric feature fusion strategy \cite{HI2D2tgrs}, intuitive assimilation modality-driven crossmodal subspace clustering method \cite{IamCSCtgrs}, and bilinear fusion \cite{wm23tgrs} have been adopted. For classifier design, logistic regression and Multi-Layer Perceptron (MLP) are commonly employed.

\begin{figure*}[ht]
\centering
\includegraphics [width=5.5in]{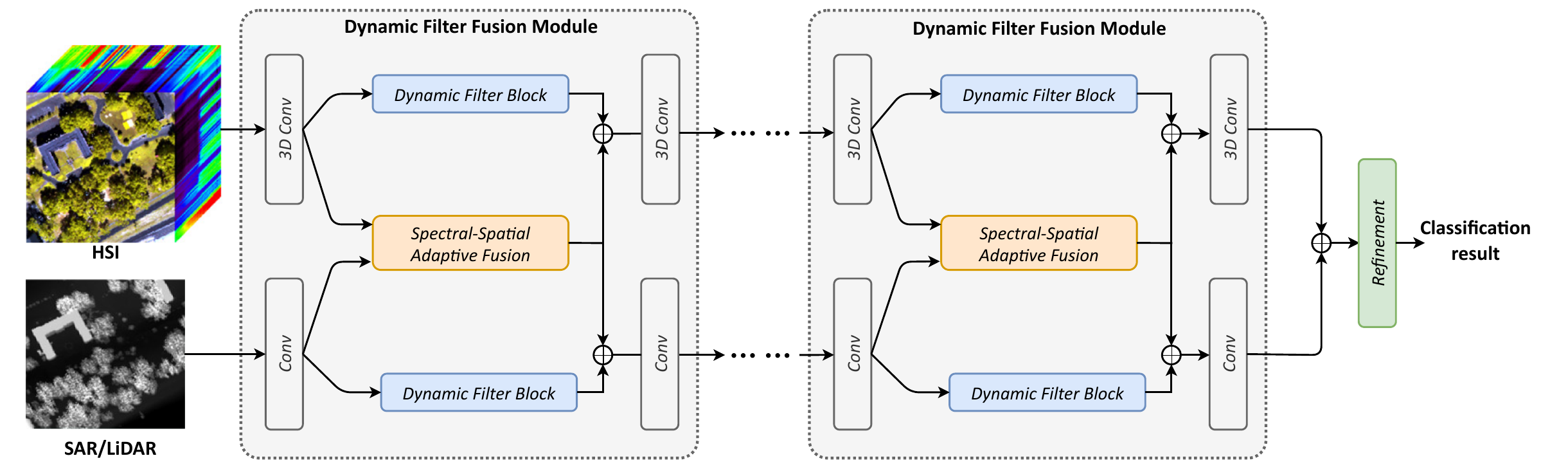}
\caption{Overview of the proposed Dynamic Frequency Feature Fusion Network (DFFNet). It incorporates several Dynamic Filter Fusion Modules (DFFMs) to learn high-quality multi-source feature representations. In each DFFM, the input features are split and fed into the Dynamic Filter Block (DFB) and Spectral-Spatial Adaptive Fusion Block (SSAFB).}
\label{fig_network}
\end{figure*}

Although existing methods have demonstrated satisfactory classification performance on benchmark datasets, they still exhibit limitations in the following aspects: \emph{\textbf{1) Limited adaptability to frequency-domain characteristics of diverse input features.}} Current methods struggle to adapt to the frequency-domain characteristics of different land cover types, thereby failing to effectively capture subtle variations within the data. Developing dynamic filters for frequency-domain feature exploration is therefore essential for multi-source data classification tasks. \emph{\textbf{2) Suboptimal feature fusion strategy.}} Existing methods often fail to fully leverage the spectral richness of HSIs and the spatial/structural details of SAR/LiDAR data during feature fusion, leading to insufficient integration of complementary multi-source information.

To overcome the above limitations, we propose a \underline{\textbf{D}}ynamic \underline{\textbf{F}}requency Feature \underline{\textbf{F}}usion \underline{\textbf{Net}}work, dubbed as \textbf{DFFNet}. Specifically, with the aim of enhance dynamic modeling capabilities for frequency domain feature representation, we propose a Dynamic Filter Block (DFB). The DFB dynamically learn the filter kernels in the frequency domain by aggregating the input features, thereby injecting the frequency contextual knowledge into every weight of the filter kernels. Additionally, in order to enhance the feature interactions between HSI and SAR/LiDAR data, we propose Spectral-Spatial Adaptive Fusion Block (SSAFB) for the purpose of feature fusion. SSAFB enhances the spectral and spatial attention weight interactions via channel shuffle operation. It achieves a significant enhancement in the classification performance. Extensive experiments conducted on two datasets involving HSI and SAR/LiDAR have thoroughly validated that our DFFNet outperforms other state-of-the-art methods.

Our main contributions are summarized as follows:

\begin{itemize}

\item We propose the DFB to enhance the dynamic modeling capabilities for frequency domain feature representation. It effectively capture subtle frequency variations among different land cover types. 

\item We present the SSAFB for spectral and spatial feature interaction and fusion. It enhances the spectral and spatial attention weight interactions via channel shuffle, thereby providing comprehensive cross-modal feature fusion.

\end{itemize}

\section{Methodology}

The framework of the proposed DFFNet is depicted in Fig. \ref{fig_network}. The DFFNet consists of two components:  1) The primary network incorporates two Dynamic Filter Fusion Modules (DFFMs) to learn high-quality multi-source feature representations. 2) A refinement module comprising two fully-connected layers is designed for land cover classification.

To reduce the computational burden and the redundancy in spectral information, principal component analysis (PCA) is first used to reduce the dimensionality of the spectral dimension in HSI. Let $\mathbf{I}^H$ denote the HSI after PCA, and $\mathbf{I}^X$ denote the input SAR/LiDAR data. First, 3D convolution is employed for $\mathbf{I}^H$ preprocessing, and 2D convolution is used for $\mathbf{I}^X$ preprocessing. Subsequently, multi-source features are fed into the DFFM for feature extraction and fusion. The DFFM is iterated two times, and the extracted features are further processed by convolution layers to generate the output.

In the DFFM, the input features are split and fed into the Dynamic Filter Block (DFB) and Spectral-Spatial Adaptive Fusion Block (SSAFB), respectively. The DFB is designed to exploit frequency domain features, while the SSAFB is intended to integrate spatial and channel-wise features for improving cross-modal feature fusion. In the subsequent subsections, we will elaborate on the DFB and SSAFB in detail.

\begin{figure}[]
    \centering
    \includegraphics[width=\linewidth]{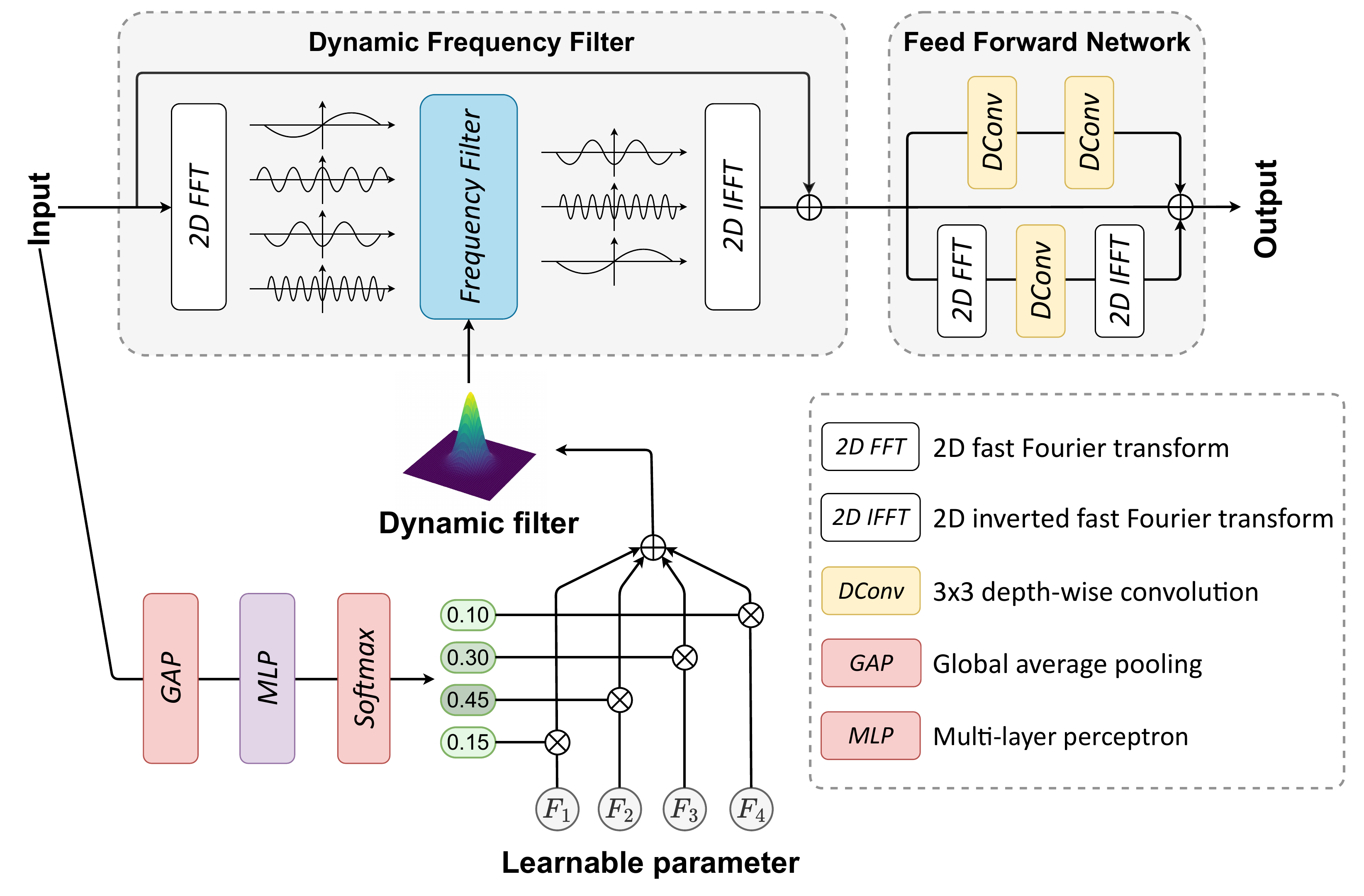}
    \caption{Details of the Dynamic Filter Block (DFB). It dynamically learn a set of filter kernels by aggregating the input features. These filters capture diverse frequency domain characteristics of different land cover types.}
    \label{fig_dfb}
\end{figure}

\subsection{Dynamic Filter Block}

Feature extraction is critical for multi-source data classification. In FCNet \cite{fcnet25tgrs}, global filter in the frequency domain are employed to model long-range dependencies and semantic relationships across the entire image. However, it struggle to adapt to the frequency-domain characteristics of different land cover types. To solve the problem, we design the Dynamic Filter Block (DFB) which dynamically learn a set of filter kernels by aggregating the input features. Details of the DFB are depicted in Fig. \ref{fig_dfb}. For the input feature $\mathbf{X}_{in}$, we first employ the Fast Fourier Transform (FFT) to the input feature $\mathbf{X}_{in}$, and converts the input signals from the spatial domain to the frequency domain. The frequency domain feature $f(u,v)$ is computed as follows:
\begin{equation}
    f(u,v)=\sum^{H-1}_{y=0}\sum^{W-1}_{x=0}\mathbf{X}_{in}(x,y)e^{-j2\pi(\frac{ux}{W}+\frac{vy}{H})},
\end{equation}
where $\mathbf{X}_{in}(x, y)$ represents the input feature at spatial position $(x,y)$. $f(u,v)$ denotes the frequency domain feature at frequency coordinates $(u, v)$, where $u$ and $v$ are the frequency indices in the horizontal and vertical directions, respectively. $j$ is the imaginary unit. 

In the frequency domain, features are represented as fluctuations of different frequencies, providing an opportunity to identify meaningful patterns. Recently, dynamic neural networks whose structure or weights can change during runtime based on the input data, often achieve superior results in challenging domains \cite{dynamic24aaai}. Inspired by this, we design dynamic filters that adaptively focus on distinct components (edges and textures) within HSI and SAR/LiDAR data according to the input features. This approach enables enhanced classification performance for a wide range of land cover types. We filter the frequency domain features with the aim of eliminating the irrelevant frequency components as follows:

\begin{equation}
    \hat{f}(u,v)=\mathcal{K}(\mathbf{X}_{in}) \odot f(u,v),
\end{equation}
where $\odot$ is the element-wise product, and $\mathcal{K}$ denotes the function that generates the dynamic filter. After that, we employ an Inverse Fast Fourier Transform (IFFT) to convert the features from the frequency domain to the spatial domain:

\begin{equation}
\mathbf{X}_{out}=\frac{1}{HW}\sum^{H-1}_{v=0}\sum^{W-1}_{u=0}\hat{f}(u,v)e^{j2\pi(\frac{ux}{W}+\frac{vy}{H})}.
\end{equation}

\textbf{Dynamic Filter Generation.} To generate the dynamic filter, a set of learnable parameters $\mathbf{F}$ are used as the filter basis so that $\mathbf{F}=\{F_1, F_2, \ldots, F_N\}$. $\mathbf{F}$ is initialized randomly, and has the same size as the filter $f(u,v)$. To generate the dynamic filter according to the input data, we initially conduct Global Average Pooling (GAP) on the input feature $\mathbf{X}_{in}$. Next, MLP is employed to generate weights for $\mathbf{F}$, as illustrated in Fig. \ref{fig_dfb}. Softmax is utilized for weights normalization of each filter. The output is computed as follows:

\begin{equation}
    \mathcal{K}(\mathbf{X}_{in})=\textrm{Softmax}(\textrm{MLP}( \textrm{Pool} (\mathbf{X}_{in}))) \otimes \mathbf{F},
\end{equation}
where $\otimes$ denotes the tensor multiplication, and Pool denotes the down-sampling operation using GAP.

\textbf{Feed Forward Network.} As shown in Fig. \ref{fig_dfb}, the obtained feature $\mathbf{X}_{out}$ is fed into the Feed Forward Network (FFN) for non-linear transformation. The FFN is composed of an identity branch, a spatial convolutional branch, and a frequency convolutional branch. Both the spatial and frequency convolutional branches each contain depth-wise convolution. The distinction lies in that one conducts convolution in the spatial domain, while the other carries out convolution in the frequency domain. Therefore, the complementary information between spatial and frequency domain features are exploited, and the non-linear feature transformation is effectively enhanced. 

\subsection{Spectral-Spatial Adaptive Fusion Block}

\begin{figure}
  \centering
  \includegraphics[width=3.4in]{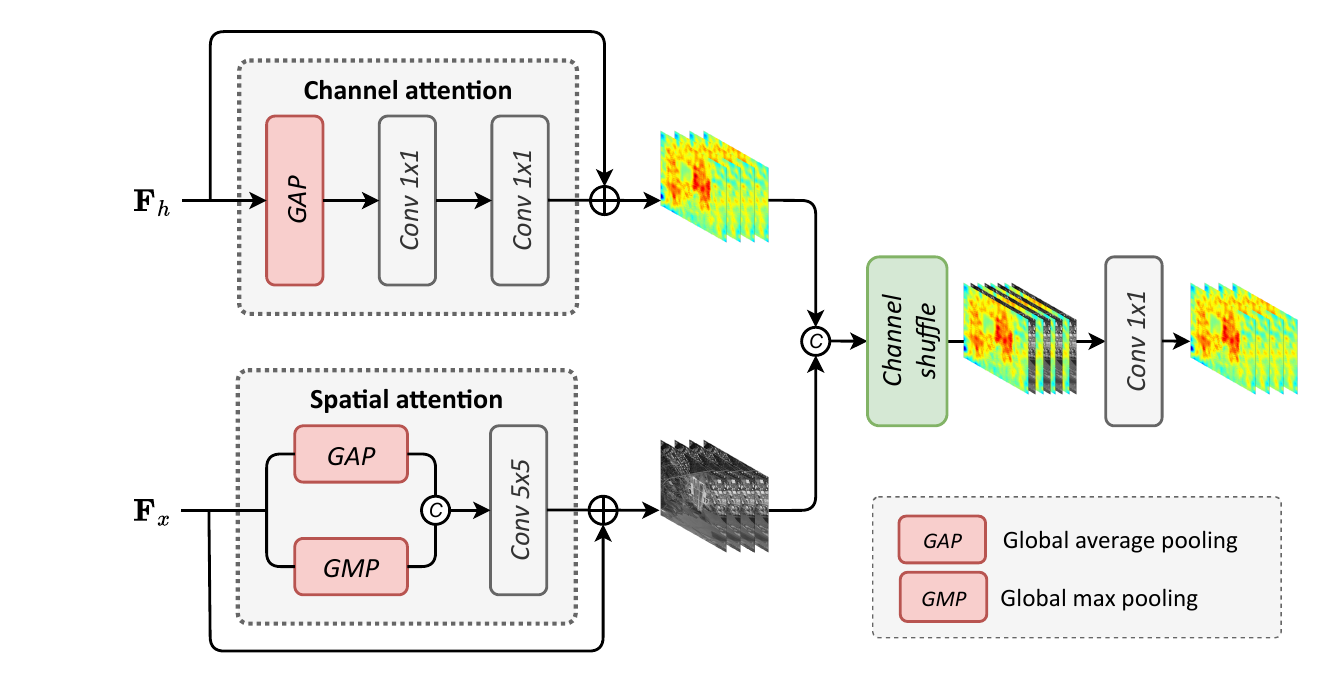}
  \caption{Illustration of the Spectral-Spatial Adaptive Fusion Block (SSAFB). It enhances the spectral and spatial attention weight interactions via channel shuffle, thereby providing comprehensive cross-modal feature fusion.}
  \label{fig_ssafb}
\end{figure}

Existing methods fail to fully exploit the advantages of spectral features in HSI and spatial features in SAR/LiDAR data during feature fusion, resulting in suboptimal integration of the complementary multi-modal information. We argue that the interactions between spectral and spatial attention weights need to be enhanced. To this end, we design the Spectral-Spatial Adaptive Fusion Block (SSAFB) which integrates the spectral information of HSI and the spatial features of SAR/LiDAR. As shown in Fig. \ref{fig_ssafb}, the features from HSI is denoted as $\mathbf{F}_h$ and the features from SAR/LiDAR is denoted as $\mathbf{F}_x$. $\mathbf{F}_h$ is first enhanced by channel attention, and $\mathbf{F}_x$ is enhanced by spatial attention as follows:

\begin{equation}
    \mathbf{F}'_h=\mathbf{F}_h+ \mathcal{C}_{1\times1}(\sigma \mathcal{C}_{1\times1}(\textrm{GAP}(\mathbf{F}_h))),
\end{equation}

\begin{equation}
    \mathbf{F}'_x=\mathbf{F}_x+\mathcal{C}_{5\times5}([ \mathrm{GAP}(\mathbf{F}_x), \mathrm{GMP}(\mathbf{F}_x)]),
\end{equation}
where $\sigma$ denotes the ReLU activation function, $\mathcal{C}_{k\times k}(\cdot)$ denotes convolution with a $k\times k$ kernel. In channel attention, GAP is the global average pooling across the channel dimension. In spatial attention, GAP and GMP denote the global average pooling and global max pooling across the spatial dimension. We fuse $\mathbf{F}'_h$ and $\mathbf{F}'_x$ together via concatenation. To effectively exchange the information of HSI and SAR/LiDAR data, every channel of $\mathbf{F}'_h$ and $\mathbf{F}'_x$ are rearranged in an alternating manner with channel shuffle operation. Next, a $1\times1$ convolution layer is used to reduce the channel size. The output is computed as follows:

\begin{equation}
    \mathbf{F}_o=\mathcal{C}_{1\times1}(\mathrm{CS}([\mathbf{F}'_h, \mathbf{F}'_x])),
\end{equation}
where $\mathrm{CS}(\cdot)$ is the channel shuffle operation. The output $\mathbf{F}_o$ has the same size with the input $\mathbf{F}_h$ and $\mathbf{F}_x$. As shown in Fig. \ref{fig_network}, in the the $\mathbf{F}_o$ are fused with the HSI feature and SAR/LiDAR feature again by element-wise summation. The SSAFB enhances the spectral and spatial attention weight interactions via channel shuffle operation. Therefore, more meaningful information is emphasized to effectively improve the land cover classification performance.

\begin{figure*}[]
  \centering
  \includegraphics[width=5.0in]{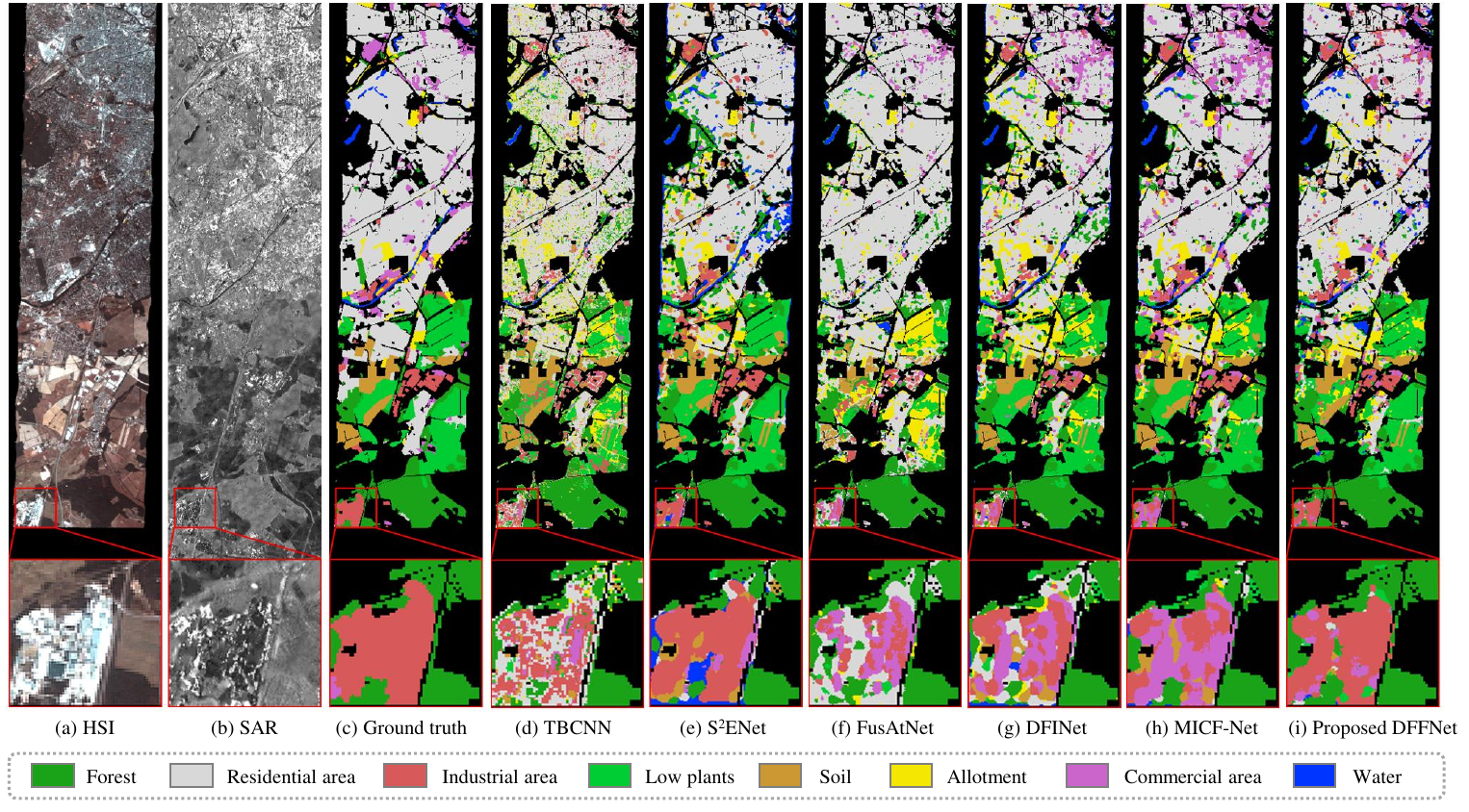}
  \caption{Classification results of different methods on the Berlin dataset.}
  \label{fig_berlin}
\end{figure*}

\section{Experimental Results and Analysis}

\subsection{Datasets and Experimental Setting}

We evaluate the performance of the proposed DFFNet on two multi-source remote sensing datasets for land cover classification. Specifically, the first dataset is the Berlin dataset. The dataset consists of HSI and SAR data. It contains 8 classes for ground truth labels, and consists of 1723$\times$476 pixels. The second dataset is the Houston2013 dataset, which consists of HSI and LiDAR data. It is part of the 2013 IEEE GRSS Data Fusion Contest, provides a unique perspective on urban land cover in Houston. The dataset includes HSI with 144 spectral bands covering wavelengths from 380 to 1050 nm, along with LiDAR data that provides precise elevation information. The Houston2013 dataset includes 15 distinct land cover types, offering valuable insights for remote sensing and urban analysis.

The proposed DFFNet was conducted on NVIDIA GeForce RTX 4090 GPU. The training phase spanned over 100 epochs. The Adam optimizer is used with the learning rate of 0.0001. The batch size is set as 128. To evaluate the performance of the proposed DFFNet, we compare it against six state-of-the-art methods: TBCNN \cite{tbcnn}, S$^2$ENet \cite{fs22grsl}, FusAtNet \cite{fusatnet}, DFINet \cite{gyh22tgrs}, and MICF-Net \cite{micfnet}. These methods are assessed through both visual comparisons and quantitative metrics, including Overall Accuracy (OA), Average Accuracy (AA), and Kappa coefficient. OA represents the proportion of correctly classified pixels relative to the total number of pixels in the dataset. AA provides a balanced assessment by considering the accuracy for each class. The Kappa coefficient offers a more robust evaluation of accuracy by accounting for the potential agreement that could occur by chance.

\begin{table}[]
\centering
\caption{Experimental results on the Berlin dataset.}
\label{table_berlin}
\resizebox{0.98\linewidth}{!}{
\begin{tabular}{c|ccccccc}
\hline\toprule
Class  & TBCNN & S$^2$ENet & FusAtNet & DFINet & MICF-Net & DFFNet \\ 
\midrule
Forest           & 81.75 & 83.27 & \textbf{86.24} & 82.04 & 84.71 & 82.45 \\
Residential area & 76.26 & 72.07 & \textbf{91.38} & 77.78 & 76.19 & 79.59 \\
Industrial area  & 39.67 & 46.66 & 19.76 & 47.96 & 44.15 &  \textbf{54.69} \\
Low plants       & 49.78 & 72.08 & 20.00 & 77.78 & 77.96 & \textbf{81.88} \\
Soil             & \textbf{89.42} & 77.94 & 48.72 & 87.85 & 67.13 & 75.84 \\
Allotment        & 54.36 & \textbf{70.62} & 38.89 & 44.75 & 61.95 & 66.87 \\
Commercial area  & 4.65  & \textbf{36.48} & 18.47 & 29.85 & 29.28 & 25.29 \\
Water            & 41.93 & 54.64 & 29.61 & \textbf{60.09} & 56.98 & 52.18 \\
\midrule
OA               & 67.60 & 70.38 & 70.91 & 73.69 & 72.59 & \textbf{75.42} \\
AA               & 54.72 & 64.22 & 44.13 & 63.51 & 62.39 & \textbf{64.85} \\
Kappa            & 50.96 & 57.73 & 51.07 & 61.02 & 59.72 & \textbf{63.22} \\ 
\bottomrule\hline
\end{tabular}}
\end{table}

\subsection{Parameter Analysis}

\begin{figure}[htb]
\centering
\includegraphics[width=\linewidth]{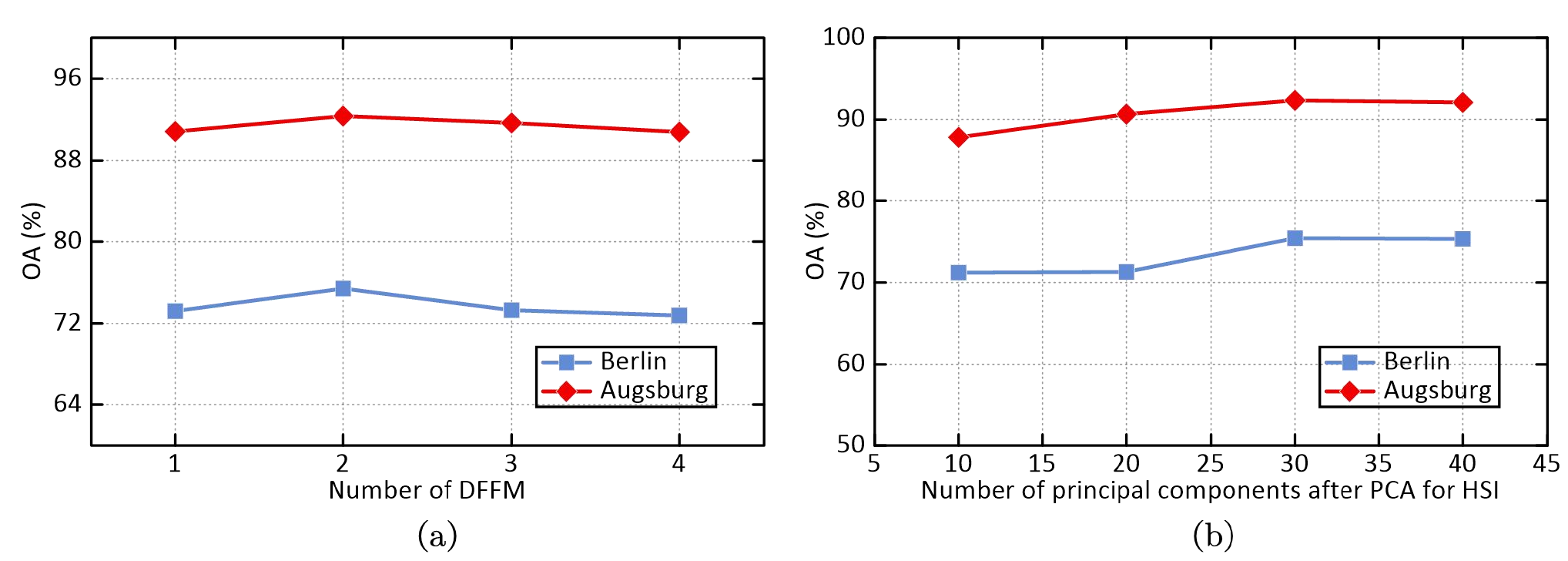}
\caption{Parameter analysis. (a) The relationship between OA and the number of DFFM. (b) The relationship between OA and the number of principal components after PCA for HSI.}
\label{fig_para}
\end{figure}

\textbf{Number of DFFM.} The number of DFFM is an important parameter that may affect the classification performance. We test different number of DFFN from 1 to 4. The experimental results is shown in Fig. \ref{fig_para}(a). When the number of DFFM is set to 2, our DFFNet achieves the best performance on both dataset. Therefore, in our following experiments, we use two DFFM in our network.

\textbf{The Number of Principal Components for HSI.} We tested the number of principal components after PCA for HSI. The experimental results is shown in Fig. \ref{fig_para}(b). It can be observed that when the number of principal components is set to 30, our DFFNet achieves the best classification performance.

\subsection{Experimental Results and Discussion}

\textbf{Results on the Berlin Dataset.} The classification results on the Berlin dataset is shown in Fig. \ref{fig_berlin}, and the corresponding quantitative evaluations are illustrated in Table \ref{table_berlin}. The OA value of the proposed DFFNet achieves 75.42\%, which is significantly higher than the OA values of the other methods. Especially, the proposed DFFNet achieves notable improvements in Industrial Area and Low Plants. It is evident that our DFFNet excels at handling challenging classes and produces results with smooth boundaries. Therefore, we draw the conclusion that our DFFNet outperforms the other methods on the Berlin dataset. 

\textbf{Results on the Houston2013 Dataset.} The quantitative evaluations of different methods on the Houston2013 dataset are illustrated in Table \ref{table_houston}. It can be observed that the OA value for the proposed DFFNet achieves 92.35\%. It surpasses the other methods by at least 2.79\%. It indicates that DFFNet effectively fuse cross-modal data features, providing a more comprehensive feature representation for multi-source data classification. Specifically, the TBCNN uses two-branch CNNs for feature extraction. Our DFFNet outperforms TBCNN since the dynamic filters in DFB emphasize important components in the frequency domain. Additionally, DFFNet performs better than the FusAtNet in which cross-attention fusion is are employed. It demonstrated that SSAFB in our DFFNet is more efficient in cross-modal feature fusion, since spectral and spatial attention interaction is enhanced via channel shuffle operation.

\begin{table}[]
\centering
\caption{Experimental results on the Houston2013 dataset.}
\label{table_houston}
\resizebox{0.98\linewidth}{!}{
\begin{tabular}{c|cccccc}
\hline\toprule
Class  & TBCNN & S$^2$ENet & FusAtNet & DFINet & MICF-Net & DFFNet \\ 
\midrule
Health grass    & 82.72 & 81.48 & 82.53 & 82.91 & 82.90 & \textbf{83.10} \\
Stressed grass  & 83.18 & 84.68 & 85.15 & 85.15 & 83.46 & \textbf{99.72} \\
Synthetic grass & \textbf{100.0} & 99.01 & 98.81 & 99.80 & 99.60 & 98.61 \\
Trees & 94.03 & 92.14 & 92.90 & 92.33 & 95.63 & \textbf{96.88} \\
Soil & 99.15 & \textbf{100.0} & \textbf{100.0} & 99.81 & \textbf{100.0} & 99.71 \\
Water      & 99.30 & \textbf{100.0} & 98.60 & \textbf{100.0} & 95.80 & 95.80 \\
Residential & 79.66 & 89.93 & 85.45 & \textbf{93.47} & 83.49 & 79.20 \\
Commercial & 55.27 & 90.18 & 81.81 & 82.58 & 81.91 & \textbf{93.46} \\
Road & 75.83 & \textbf{92.63} & 83.76 & 89.05 & 82.15 & 87.54 \\
Highway & 62.55 & 64.86 & 53.67 & 56.56 & 66.22 & \textbf{84.27} \\
Railway & 96.87 & \textbf{97.60} & 79.65 & 94.15 & 95.59 & 94.82 \\
Parking lot1 & 86.55 & 88.09 & 91.26 & 94.72 & 94.62 & \textbf{96.16} \\
Parking lot2 & 53.68 & 91.93 & 81.05 & 89.12 & 81.40 & \textbf{92.98} \\
Tennis court   & 98.79 & \textbf{100.0} & \textbf{100.0} & \textbf{100.0} & \textbf{100.0} & \textbf{100.0} \\
Running track    & 98.10 & \textbf{100.0} & 98.73 & \textbf{100.0} & 99.79 & \textbf{100.0} \\
\midrule
OA & 82.91 & 89.56 & 85.32 & 88.59 & 88.09 & \textbf{92.35} \\
AA & 84.38 & 91.50 & 87.56 & 90.63 & 89.65 & \textbf{93.48} \\
Kappa & 81.43 & 88.69 & 84.12 & 87.64 & 87.12 & \textbf{91.70} \\ 
\bottomrule\hline
\end{tabular}}
\end{table}

\textbf{Computational Cost Analysis.} We present the  Floating Point Operations Per Seconds (FLOPs), model parameters, and inference time of different methods for computational cost analysis in Table \ref{table_complex}. It can be noted that our DFFNet exhibits excellent trade-off in terms of model parameters, computational load, and inference time. The number of parameters of DFFNet amount to 1.2829M, which is at a moderate level. As for inference time, DFFNet takes 0.2387 seconds, showing a clear advantage over most of the comparative methods.

\begin{table}[ht]

\centering
\caption{Computational costs of different methods.}
\label{table_complex}
\resizebox{0.95\linewidth}{!}{
\begin{tabular}{c|ccccccc}
\hline\toprule
& TBCNN & S$^2$ENet & FusAtNet & DFINet & MICF-Net & DFFNet\\ 
\midrule
Params (M) & 0.3875 & 1.4701 & 37.7177 & 1.3155 & 2.1375 & 1.2829\\
FLOPs (G) & 0.0063 & 0.1778 & 3.5619 & 0.1191 & 0.0536 & 0.0303 \\
Inference time (s) & 0.2236 & 0.2489 & 0.2469 & 0.3182 & 0.2965 & 0.2387 \\
\bottomrule\hline
\end{tabular}}
\end{table}

\begin{table}[ht]
\centering
\caption{Ablation study of the proposed DFFNet.}
\label{table_ablation}
\resizebox{0.65\linewidth}{!}{
\begin{tabular}{c|ccc} 
\toprule
\multirow{2}{*}{Method}
    & \multicolumn{2}{c}{OA on different datasets ($\%$)} \\ \cmidrule{2-3}
& Houston2013 & Berlin\\ 
\midrule
w/o SSAFB and DFB & 87.51 & 72.29\\
w/o SSAFB   & 90.81  & 74.53 \\  
w/o DFB   & 89.95  & 73.32 \\ 
Proposed DFFNet & \textbf{92.35}  & \textbf{75.42}  \\  
\bottomrule
\end{tabular}}
\end{table}

\subsection{Ablation Study}

We carry out a series of ablation experiments on two datasets. We individually eliminate the SSAFB (referred as ``w/o SSAFB") and remove the DFB (referred as ``w/o DFB"). The experimental results are presented in Table \ref{table_ablation}. It can be observed that our DFFNet consistently attains superior performance compared to its variants on both datasets. This clearly illustrates the indispensability of the SSAFB and DFB incorporated within the DFFNet.

\section{Conclusions}

In this letter, we devised the Dynamic Frequency Feature Fusion Network (DFFNet) for HSI and SAR/LiDAR data joint classification. The DFB is designed to dynamically learn the filter kernels in the frequency domain by aggregating the input features. The frequency contextual knowledge is injected into frequency filter kernels. Additionally, SSFAB is presented for cross-modal feature fusion. It enhances the spectral and spatial attention weight interactions via channel shuffle operation, thereby providing comprehensive cross-modal feature fusion. Experiments on two benchmark datasets demonstrated that DFFNet achieved clearly superior performance compared to state-of-the-art methods.

\bibliographystyle{IEEEtran}
\bibliography{re} 

\begin{thebibliography}{10}
\providecommand{\url}[1]{#1}
\csname url@samestyle\endcsname
\providecommand{\newblock}{\relax}
\providecommand{\bibinfo}[2]{#2}
\providecommand{\BIBentrySTDinterwordspacing}{\spaceskip=0pt\relax}
\providecommand{\BIBentryALTinterwordstretchfactor}{4}
\providecommand{\BIBentryALTinterwordspacing}{\spaceskip=\fontdimen2\font plus
\BIBentryALTinterwordstretchfactor\fontdimen3\font minus
  \fontdimen4\font\relax}
\providecommand{\BIBforeignlanguage}[2]{{%
\expandafter\ifx\csname l@#1\endcsname\relax
\typeout{** WARNING: IEEEtran.bst: No hyphenation pattern has been}%
\typeout{** loaded for the language `#1'. Using the pattern for}%
\typeout{** the default language instead.}%
\else
\language=\csname l@#1\endcsname
\fi
#2}}
\providecommand{\BIBdecl}{\relax}
\BIBdecl

\bibitem{luo24tgrs}
F.~Luo, T.~Zhou, J.~Liu, T.~Guo, X.~Gong, and X.~Gao, ``{DCENet}: Diff-feature
  contrast enhancement network for semi-supervised hyperspectral change
  detection,'' \emph{IEEE Transactions on Geoscience and Remote Sensing},
  vol.~62, pp. 1--14, 2024.

\bibitem{gyh24tip}
Y.~Gao, W.~Li, J.~Wang, M.~Zhang, and R.~Tao, ``Relationship learning from
  multisource images via spatial-spectral perception network,'' \emph{IEEE
  Transactions on Image Processing}, vol.~33, pp. 3271--3284, 2024.

\bibitem{nk25grsl}
K.~Ni, Z.~Li, C.~Yuan, Z.~Zheng, and P.~Wang, ``Selective spectral–spatial
  aggregation transformer for hyperspectral and lidar classification,''
  \emph{IEEE Geoscience and Remote Sensing Letters}, vol.~22, pp. 1--5, 2025.

\bibitem{ghm23tgrs}
H.~Gao, H.~Feng, Y.~Zhang, S.~Xu, and B.~Zhang, ``Amsse-net: Adaptive
  multiscale spatial–spectral enhancement network for classification of
  hyperspectral and lidar data,'' \emph{IEEE Transactions on Geoscience and
  Remote Sensing}, vol.~61, pp. 1--17, 2023.

\bibitem{HI2D2tgrs}
H.~H. Y. S. G.~S. Wenbo~Yu, Lianru~Gao, ``Hi2d2fnet: Hyperspectral intrinsic
  image decomposition guided data fusion network for hyperspectral and lidar
  classification,'' \emph{IEEE Transactions on Geoscience and Remote Sensing},
  vol.~61, pp. 1--15, 2023.

\bibitem{IamCSCtgrs}
Y.~S. L. S. C.~Z. W.~Yu, H.~Huang and G.~Shen, ``Iamcsc: Intuitive assimilation
  modality driven crossmodal subspace clustering for land-cover identification
  and hyperspectral-lidar fusion,'' \emph{IEEE Transactions on Geoscience and
  Remote Sensing}, vol.~63, pp. 1--13, 2025.

\bibitem{wm23tgrs}
M.~Wang, F.~Gao, J.~Dong, H.-C. Li, and Q.~Du, ``Nearest neighbor-based
  contrastive learning for hyperspectral and lidar data classification,''
  \emph{IEEE Transactions on Geoscience and Remote Sensing}, vol.~61, pp.
  1--16, 2023.

\bibitem{fcnet25tgrs}
J.~Zhang, F.~Gao, Y.~Gan, J.~Dong, and Q.~Du, ``Frequency-compensated network
  for daily {Arctic} sea ice concentration prediction,'' \emph{IEEE
  Transactions on Geoscience and Remote Sensing}, vol.~63, pp. 1--15, 2025.

\bibitem{dynamic24aaai}
Y.~Tatsunami and M.~Taki, ``{FFT}-based dynamic token mixer for vision,'' in
  \emph{Proceedings of the AAAI Conference on Artificial Intelligence}, 2024,
  pp. 15\,328--15\,336.

\bibitem{tbcnn}
X.~Xu, W.~Li, Q.~Ran, Q.~Du, L.~Gao, and B.~Zhang, ``Multisource remote sensing
  data classification based on convolutional neural network,'' \emph{IEEE
  Transactions on Geoscience and Remote Sensing}, vol.~56, no.~2, pp. 937--949,
  2018.

\bibitem{fs22grsl}
S.~Fang, K.~Li, and Z.~Li, ``{S2ENet}: Spatial–spectral cross-modal
  enhancement network for classification of hyperspectral and {LiDAR} data,''
  \emph{IEEE Geoscience and Remote Sensing Letters}, vol.~19, pp. 1--5, 2022.

\bibitem{fusatnet}
S.~Mohla, S.~Pande, B.~Banerjee, and S.~Chaudhuri, ``{FusAtNet}: Dual attention
  based spectrospatial multimodal fusion network for hyperspectral and {LiDAR}
  classification,'' in \emph{IEEE Conference on Computer Vision and Pattern
  Recognition Workshops (CVPRW)}, 2020, pp. 416--425.

\bibitem{gyh22tgrs}
Y.~Gao, W.~Li, M.~Zhang, J.~Wang, W.~Sun, R.~Tao, and Q.~Du, ``Hyperspectral
  and multispectral classification for coastal wetland using depthwise feature
  interaction network,'' \emph{IEEE Transactions on Geoscience and Remote
  Sensing}, vol.~60, pp. 1--15, 2022.

\bibitem{micfnet}
X.~Tang, Y.~Zou, J.~Ma, X.~Zhang, F.~Liu, and L.~Jiao, ``Multiple information
  collaborative fusion network for joint classification of hyperspectral and
  {LiDAR} data,'' \emph{IEEE Transactions on Geoscience and Remote Sensing},
  vol.~62, pp. 1--16, 2024.

\end{thebibliography}

\end{document}